\documentclass[pre,aps,twocolumn,showpacs]{revtex4}
\usepackage{graphicx} 
\usepackage{dcolumn} 
\usepackage{bm} 

\begin{document}

\title{Energy levels of a parabolically confined quantum dot \\
in the presence of spin-orbit interaction}

\author{W. H. Kuan}
\address{Department of Electrophysics, National Chiao Tung
University, Hsinchu 30010, Taiwan}
\author{C. S. Tang}
\address{Physics Division, National Center for Theoretical Sciences,
P.O. Box 2-131, Hsinchu 30013, Taiwan}
\author{W. Xu}
\address{Department of Theoretical Physics, Research School of Physical
Sciences and Engineering, Australian National University,
Canberra, ACT 0200, Australia}


\begin{abstract}
We present a theoretical study of the energy levels in a
parabolically confined quantum dot in the presence of the Rashba
spin-orbit interaction (SOI). The features of some low-lying
states in various strengths of the SOI are examined at finite
magnetic fields. The presence of a magnetic field enhances the
possibility of the spin polarization and the SOI leads to
different energy dependence on magnetic fields applied.
Furthermore, in high magnetic fields, the spectra of low-lying
states show basic features of Fock-Darwin levels as well as Landau
levels.
\end{abstract}


\date{\today} \maketitle \clearpage

\section{introduction}

The state-of-the-art material engineering and nano-fabrication
techniques have made it possible to realize advanced semiconductor
devices at atomic scales, such as quantum dots in which the
electron motion along all directions is quantized and conducting
electrons are confined within the nanometer distances. In such a
system, few electrons are confined within a point-like structure
so that it can behave as an artificial atom and, consequently, be
used as electronic and optical devices such as memory chip
\cite{Thornton}, quantum computer
\cite{Loss,Benjamin,Lent,Tougaw}, quantum cryptography
\cite{Molotkov}, quantum-dot laser \cite{Saito}, etc. In recent
years, the electronic, transport, optical and optoelectronic
properties of spin-degenerate quantum dots have been intensively
investigated.

On the other hand, the progress made in realizing spin polarized
electronic systems on the basis of diluted magnetic semiconductors
and narrow-gap semiconductor nanostructures has opened up a field
of spin-electronics (or spintronics). As has been pointed out in a
good review edited by Wolf and Awaschalom \cite{Wolf}, due to
unique nature of the SOI in electronic materials, quantum
transport of electrons in a spin polarized system differs
fundamentally from that in a spin-degenerate device. Thus, the
spin-dependent effects can offer new mechanisms and schemes for
information storage and for increasing the speed of data
processing. As a result, the electronic devices, such as
spin-transistor \cite{FET}, spin-waveguide \cite{wang},
spin-filter \cite{Tko}, etc., have been proposed. Moreover,
optical methods for injection, modulation and detection of spin
polarized electrons will eventually become the target for the
development of spin polarized nano-optoelectronic devices. At
present, one of the major challenges for the application of the
spintronic systems as working devices is to optimize the spin life
times of the carriers in the devices. It has been realized that in
contrast to the diluted magnetic semiconductors in which the SOI
is induced by the presence of an external magnetic field, spin
splitting of carriers can be achieved in narrow-gap semiconductor
nanostructures even in the absence of the magnetic field.
Experimental data have indicated that in narrow-gap semiconductor
based quantum wells, such as in InAlAs/InGaAs heterostructures,
the higher-than-usual zero-magnetic-field spin splitting (or
spontaneous spin splitting) can be realized by the inversion
asymmetry of the microscopic confining potential due to the
presence of the heterojunction \cite{scha}. This kind of inversion
asymmetry corresponds to an inhomogeneous surface electric field
and, hence, this kind of spin-splitting is electrically equivalent
to the Rashba spin-splitting or Rashba effect \cite{Rashba}. From
the fact that InGaAs-based quantum dots are normally fabricated
using InAlAs/InGaAs heterojunctions, one would expect that the
Rashba spin splitting can be observed in these quantum dot systems
and spin polarized quantum dots can therefore be achieved.
Practically, the spin-split quantum dots can be made from, e.g.,
an InAlAs/InGaAs heterostructure with a negative bias applied to
the side gate \cite{Tarucha}.

In contrast to a spin-split quantum well structure in which the
Rashba effect can be easily identified by, e.g., the
magnetotransport experiments via measuring the Shubnikov-de Hass
(SdH) oscillations \cite{scha,Grundler}, the spintronic effects in
a quantum dot cannot be easily observed using conventional
transport measurements. Normally, optical measurements, such as
optical absorption and transmission \cite{bnm},
cyclotron-resonance \cite{Krahne}, etc., can be used to determine
the energy spectrum of a quantum dot. Although recently there are
some theoretical work published regarding the Rashba effect in
quantum dots in the presence of magnetic fields
\cite{prl,Bulgakov,Reed}, the effect of the SOI on energy spectrum
of a quantum dot has not yet been fully analyzed. It should be
noted that in Ref.~\onlinecite{prl}, the SOI was taken as a
perturbation. In order to understand how SOI affects the energy
levels of a parabolically confined quantum dot, we feel that more
theoretical work is needed and it is the prime motivation of the
present work. In this paper, we present a tractable approach to
calculate energy spectrum of an InGaAs-based quantum dots with the
inclusion of SOI induced by the Rashba effect. We would like to
examine the important and interesting consequences such as the
exchange of the energy states due to SOI and the enhancement of
the spin polarization by the presence of the magnetic fields.

In Section II the theoretical approach is developed in calculating
the electronic subband structure of a parabolically confined
quantum dot in the presence of SOI and a magnetic field. The
numerical results are presented and discussed in Section III and
the conclusions obtained from this work are summarized in Section
IV.

\section{approaches}

The device system under investigation is a typical quantum dot
formed on top of a narrow-gap semiconductor heterojunction (such
as an InGaAs/InAlAs-based heterostructure) grown along the
$Z$-direction, and the lateral confinement is formed in the
$XY$-plane. A perpendicular magnetic field $B$ is applied along
the growth-direction of the quantum dot. We consider the situation
where the SOI is mainly induced by the presence of the
InGaAs/InAlAs heterojunction due to the Rashba effect. In such a
case, the lowest order of the SOI can be obtained from, e.g., a
${\bf k}\cdot {\bf p}$ band-structure calculation \cite{scha}. It
should be noted that when the SOI is induced by the Rashba effect
due to the presence of the heterojunction and when the magnetic
field is applied along the $Z$-direction, the electronic subband
structure along the growth-direction (or the $Z$-axis) depends
very little on the SOI. Therefore, under the effective-mass
approximation, the electronic structure along the $XY$-plane and
in the $Z$-direction in a quantum dot can be treated separately.
Including the contribution from SOI, the single-electron
Hamiltonian describing the electronic system in the $XY$-plane can
be written as

\begin{equation}
H = \frac{1}{2m^\ast}(\textbf{p}-e\textbf{A})^2 +
\frac{\alpha}{\hbar}[\vec{\sigma}\times(\textbf{p}-e\textbf{A})]_z
 + V_c (r), \label{e1}
\end{equation}
where the Zeeman term is neglected for simplicity. Here, $m^\ast$
is the electron effective mass in the $XY$-plane,
$\textbf{p}=(p_X,p_Y)$ with $p_X=-i\hbar\partial\ /\partial X$ is
the momentum operator, $\textbf{A}= (Y,-X,0)B/2$ is the vector
potential induced by the magnetic field which is taken in the
symmetric gauge here for convenience, and $V_c(r)$ is the
confining potential of the quantum dot along the $XY$-plane with
$r=(X^2+Y^2)^{1/2}$. Furthermore, $\alpha$ is the Rashba parameter
which measures the strength of the SOI. Due to the Pauli spin
matrices $\vec{\sigma}=(\sigma_X,\sigma_Y,\sigma_Z)$, this
Hamiltonian is a $2\times 2$ matrix. After rewriting this
Hamiltonian in the cylindrical polar coordinate, we find that the
solution of the corresponding Schr\"odinger equation is in the
form of $\Psi(X,Y)=e^{im\theta}\psi(r)$, where $m$ is a good
quantum number and $\theta$ is an azimuthal angle to the $X$-axis.
Thus, the eigenfunction $\psi(r)$ and eigenvalue $E$ of the system
can be determined by solving
\begin{equation}
\left[\begin{array}{cc} H_0-E & \hskip 0.5truecm  e^{-i\theta} H_+ \\
-e^{i\theta} H_- & \hskip 0.5truecm H_0-E \end{array} \right]
\left[\begin{array}{cc} \psi_1(r) \\
\psi_2(r) \end{array} \right]=0, \label{e2}
\end{equation}
with
$$H_0=-\frac{\hbar^2}{2m^\ast}\Bigl(\frac{\partial^2}{\partial
r^2}+\frac{1}{r}\frac{\partial}{\partial r}\Bigr)
  -\frac{m^2}{r^2}+V^\ast(r)+\frac{m}{2}\hbar\omega_c,$$
and
$$H_\pm=\alpha  \Bigl(\frac{\partial}{\partial
r}\pm \frac{m}{r}+\frac{r}{2l^2}\Bigr).$$ Here, $\omega_c \equiv
eB/m^\ast$ is the cyclotron frequency, $l=(\hbar/eB)^{1/2}$ is the
radius of the ground cyclotron orbit, and
$V^\ast(r)=\frac{1}{8}m^\ast\omega_c^2r^2 +V_c(r)$ is the
effective trapping potential. It should be noted that the solution
of Eq. (\ref{e2}) does not depend on $\theta$, because the
eigenfunction and eigenvalue can be obtained, in principle, by
solving
\begin{equation}
[(H_0-E)^2+H_+H_-]\psi_i(r)=0, \label{e3}
\end{equation}
where $i=1,2$. Eq. (\ref{e3}) is a 4th-order differential
equation. To the best of our knowledge, there is no simple and
analytical solution to this equation. In this paper, we therefore
attempt a tractable approach to solve the problem.

For the case of a parabolically confined quantum dot, we have the
confining potential $V_c(r)=m^*\omega_0^2 r^2/2$ where $\omega_0$
is the characteristic frequency of the confinement. For
convenience, we define $\Omega = \sqrt{4\omega_0^2+\omega_c^2}$
and $a=(2\hbar/m^\ast\Omega)^{1/2}$. After setting $r^2=a x$ and
assuming $\psi_i(r)=e^{-x/2} x^{|m|/2} C_i (x)$, Eq. (\ref{e2})
readily becomes
\begin{equation}
\left[\begin{array}{cc} h_0+{\cal E} & \hskip 0.5truecm
-e^{-i\theta} h_+ \\ e^{i\theta} h_- & h_0+{\cal E} \end{array}
\right]
\left[\begin{array}{cc} C_1(x) \\
C_2(x) \end{array} \right]=0. \label{e4}
\end{equation}
Here, $h_0= x \partial^2/ \partial x^2+(|m|-x+1)\partial/
\partial x$, $h_\pm= A_\alpha\sqrt{x}\ [\partial/\partial
x+(|m|\pm m)/2x-(1-\omega_c/\Omega)/2]$, ${\cal
E}=E/\hbar\Omega-m\omega_c/2\Omega-(|m|+1)/2$, and
$A_\alpha=(\alpha/\hbar)\sqrt{2 m^*/\hbar\Omega}$. In this paper,
we are interested in finding the energy spectrum of a quantum dot
in the presence of the SOI and of a magnetic field. We assume that
the electron wavefunctions are in the form
$$C_1(x)=\sum^\infty_{N=0} C_{m,N} L^{|m|}_N(x)$$
and $$C_2(x)=\sum^\infty_{N=0} D_{m,N} L^{|m|}_N (x)$$ with
$L^m_N(x)$ being the associated Laguerre polynomial. Thus,
introducing $C_i (x)$ into Eq.~(\ref{e4}) and carrying out the
normalization, we obtain two coupled equations which determine the
energy $E$ and the coefficients $C_{m,N}$ and $D_{m,N}$
\begin{equation}
(E-E_{m,N}^0) C_{m,N}-A_\alpha \hbar\Omega \ I_{N'N}^-
D_{m+1,N'}/2=0, \label{e5}
\end{equation}
and
\begin{equation}
(E-E_{m,N}^0) D_{m,N}+A_\alpha \hbar\Omega \ I_{N'N}^+
C_{m-1,N'}/2=0, \label{e6}
\end{equation}
where $E_{m,N}^0=m\hbar\omega_c/2+(2N+|m|+1)\hbar\Omega/2$ is the
energy of a parabolically confined quantum dot in the absence of
SOI,
$$I_{N'N}^\pm = (\omega^*+1) \delta_{N',N} -(\omega^*-1)\delta_{N',N-1}
$$ when $m\geq 1$ for $I_{N'N}^-$ and $m<0$ for $I_{N'N}^+$ and,
otherwise,
$$I_{N'N}^\pm=(\omega^*-1)(N+|m|+1)\delta_{N',N} -(\omega^* +1)(N+1)
\delta_{N',N+1},$$ with
$\omega^*=\omega_c/\Omega$. These results suggest that the SOI in
a quantum dot can result in band-mixing and shifting. Using
sequence relations given by Eqs.~(\ref{e5}) and (\ref{e6}), the
electron energy $E$ as well as the coefficients $C_{m,N}$ and
$D_{m,N}$ can be determined.

\section{numerical results and discussion}

In the present study, we examine the dependence of the energy
levels and the occupation of electrons to different states on the
strength of the SOI and of the magnetic field in a few-electron
quantum dot system. We take the Rashba parameter to be $\alpha\sim
(3 - 4)\times 10^{-11}$ eV$\cdot$m, in conjunction with recent
experimental data realized in InAlAs/InGaAs heterostructures
\cite{Saito,Grundler}. The size effects of the dot are also
considered. For weakly and strongly confined dots, we take the
typical values of the confining potential as $V_c
=\hbar\omega_0=1$ meV and $10$ meV, respectively. The condition of
electron number conservation is applied to determine the Fermi
energy of the system. For demonstration of how electrons occupy to
the different energy levels in the presence of SOI, we assume
there are ten electrons in the dot. Moreover, in the present study
the effect of the SOI on lifetimes of an electron in different
states is not taken into consideration~\cite{Governale}.   The
inclusion of the many-body effects induced by Coulomb interaction
in a quantum dot is a higher order effect to affect the energy
spectrum, although the Coulomb interaction can result in an
exchange enhancement which may affect the results
qualitatively~\cite{Pfan}.  Below, we discuss the effect of the
SOI on energy levels of the low-lying states at zero and non-zero
magnetic fields. For convenience of the discussion, we label the
energy states with quantum numbers $\{N,m,\chi_s\}$, where
$\chi_s=$d or $\downarrow$ ($\chi_s=$u or $\uparrow$) referring to
the down (up) spin branch of the energy states.

\begin{figure}[b]
\includegraphics[width=0.45\textwidth]{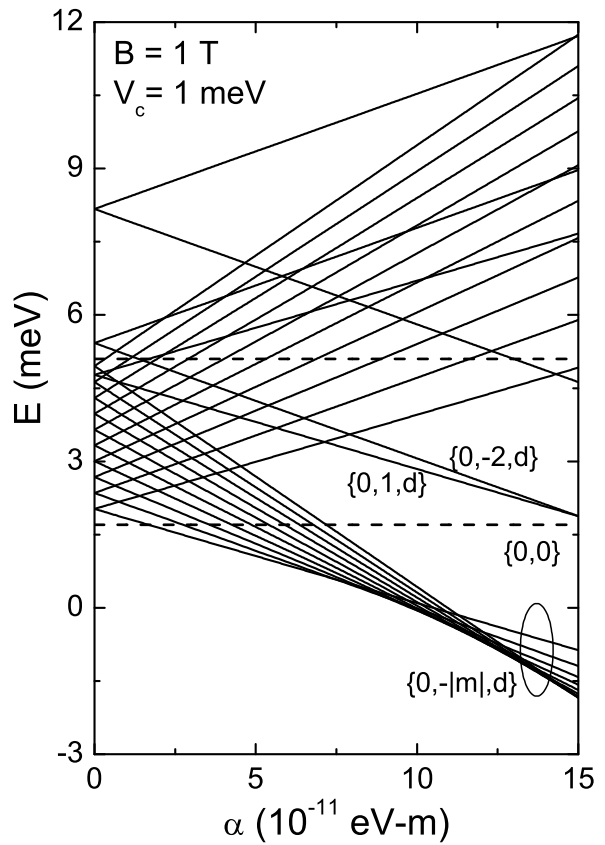}
\caption{Energy of Low-lying levels vs the Rashba parameter at B=1
T and $V_c$=1 meV. The presence of the magnetic field reduces the
spin polarization and a fully spin-polarized ten-electron dot is
achieved at $\alpha>\alpha_c\sim 7.25\times 10^{-11}$
eV$\cdot$m.}\label{fig1}
\end{figure}

\begin{figure}[bt]
\includegraphics[width=0.45\textwidth]{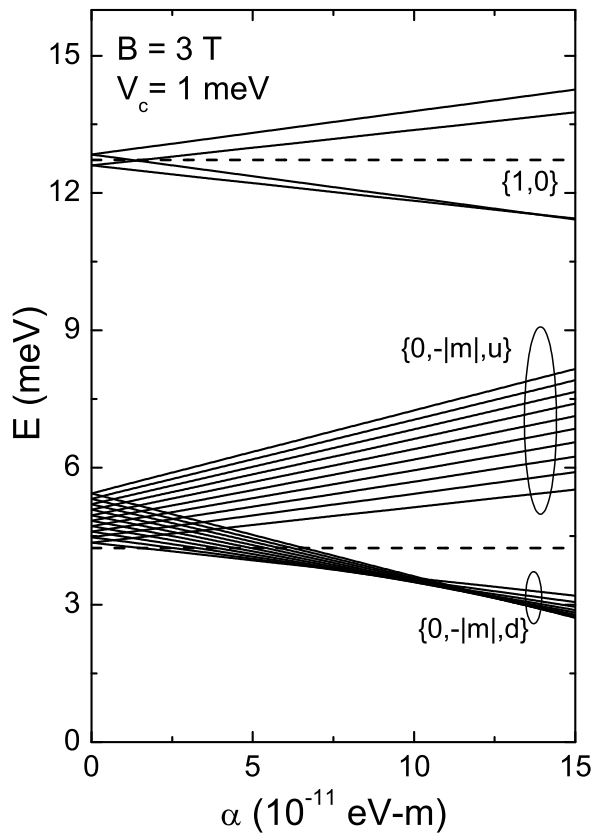}
\caption{Energy of low-lying states vs $\alpha$ at B=3 T and
$V_c$=1 meV. In the lower panel, the $\{0,-m,\downarrow\uparrow\}$
states are assembled in different groups and the dashed line
refers to the energy of the $\{0,0\}$ level. A fully
spin-polarized ten-electron dot is achieved at
$\alpha>\alpha_c\sim 6.75\times 10^{-11}$ eV$\cdot$m.}\label{fig2}
\end{figure}

When a perpendicular d.c. magnetic field is applied to a quantum
dot, the nature of spin-dependent energy spectrum can become even
richer in terms of physics due to the coupling of the magnetic
field to the confining potential of the quantum dot and to the
potential induced by SOI. It should be noted that in the present
study, in order to examine the net contribution to the energy
spectrum of a dot from the SOI, we have neglected the effects
induced by the Zeeman splitting. In Figs. \ref{fig1} and
\ref{fig2} we plot the low-lying energy levels as a function of
the Rashba parameter for different magnetic fields B = 1 T and 3
T, respectively, at a fixed quantum dot confinement $V_c=$ 1 meV.
In the absence of the Zeeman splitting, the SOI induced by the
Rashba effect at finite magnetic field can lift the magnetic
degeneracy of electrons. Namely, the states with the $\pm|m|$
number can have different energies. We find that in this
situation, the occupancy of electrons to different states has some
unique features. On one hand, the presence of the magnetic field
can increase the energy gaps among the states with the quantum
number $(N,m)$. Thus, similar to a strongly confined dot, the
magnetic field may increase the critical value $\alpha_c$ above
which the system is fully spin-polarized. This effect can be seen
at a relatively low-B-field (see Fig. \ref{fig1}) where
$\alpha_c\sim 7.25\times 10^{-11}$ eV$\cdot$m is found. On the
other hand, when the magnetic field predominates the electronic
energy spectra, the energy states with the negative magnetic
quantum number $m$ and the spin-down branches are more preferable
for an electron to stay. As a result, the presence of a B-field
can increase the spin-polarization of a quantum dot. This effect
is more pronounced in the presence of a higher magnetic field (see
Fig. \ref{fig2} where $\alpha_c\sim 6.75\times 10^{-11}$
eV$\cdot$m) so that the states with positive quantum number $m$
and spin-up orientation have higher energies relative to the
$\{0,-|m|,\downarrow\}$ states. Furthermore, after comparing the
$\alpha_c$ values obtained in Figs. \ref{fig1} and \ref{fig2}, we
note that the presence of the magnetic field can result in a lower
value of $\alpha_c$ and, consequently, the stronger
spin-polarization can be achieved in a dot in the presence of the
SOI and of a magnetic field.

\begin{figure}[bt]
\includegraphics[width=0.45\textwidth]{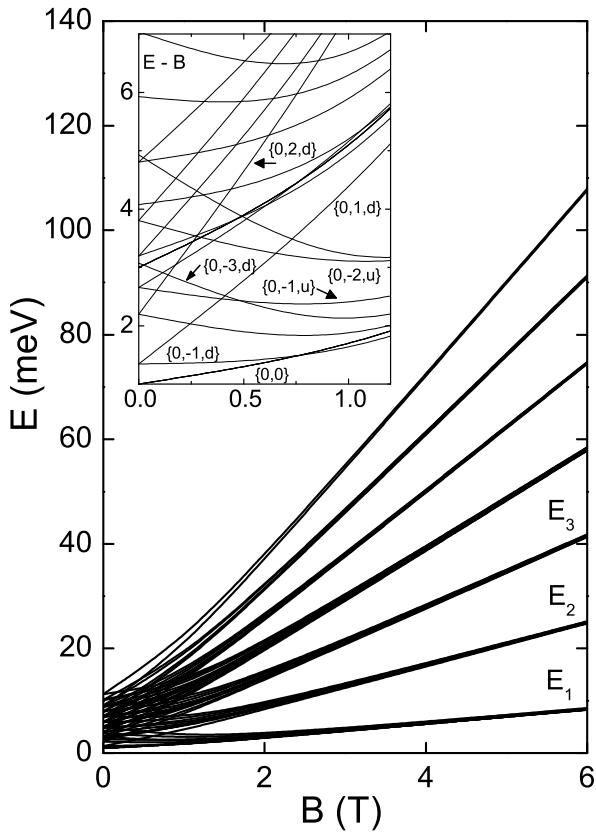}
\caption{Dependence of the energy levels on the strength of the
magnetic field at a fixed confinement $V_c$= 1 meV and a fixed
$\alpha=10^{-11}$ eV$\cdot$m. Here $E_1$, $E_2$, and $E_3$ groups
include respectively the states with
$\{0,(0,-|m|),\downarrow\uparrow\}$,
$\{1,-|m|,\downarrow\uparrow\}$ and
$\{N+m=1,N+m=1,\downarrow\uparrow\}$, and
$\{N=2,(0,-|m|),\downarrow\uparrow\}$ and
$\{N+m=2,N+m=2,\downarrow\uparrow\}$. The higher index groups can
be generated via the rule of the least sum of the quantum numbers.
The insert shows the results obtained at low-B fields.}
\label{fig3}
\end{figure}

In Figs. \ref{fig3}, we show the spectra of low-lying states as a
function of the magnetic field at a fixed Rashba parameter
$\alpha=10^{-11}$ eV$\cdot$m for a weakly confined dot $V_c = 1$
meV. We find that in such a case, because $\alpha$ is relatively
small, the electronic subband energy in the high-B field regime is
mainly determined by the strength of the magnetic field through
$E\simeq m\hbar\omega_c/2+(2N+|m|+1)\hbar\Omega$. The energy
spectra at low-B fields differ slightly from this dependence (see
the insert in Fig. \ref{fig3}). Thus, similar to a spin-degenerate
quantum dot, in the high-B field regime the energy of a state
$\{N,m,\downarrow\uparrow\}$ increases with B field and levels
with positive magnetic quantum number $m>0$ are always higher than
those with negative $m<0$. The results shown in the insert in Fig.
\ref{fig3} indicate that the cross-over of the energy states with
different $N$ numbers occurs when $B \leq 1.5$ T. When $B>2$ T,
these states can be assembled to the energy groups of E $\simeq$
14, 42, 70, 96, 124 meV, etc., corresponding to the states with
different $N$ numbers (see captions in Fig. \ref{fig3}), and these
states are affected weakly by the SOI. In the presence of very
high magnetic fields ($B \gg 2$ T in Fig. \ref{fig3}), the effects
of the SOI and the confinement of the dot can be largely
suppressed and the energy levels are therefore determined mainly
by $E_N \simeq (N+1/2)\hbar\omega_c$. Although the effect of SOI
induced by Dresselhaus splitting is not included in the present
study, these features enable us to compare qualitatively our
results with those obtained by Voskoboynikov~\cite{Vosko01}. Our
analyses give an explicit expression of energy levels and our
results clearly show the effects of SOI on energy spectra for
different sizes of quantum dots.

\begin{figure}[bt]
\includegraphics[width=0.45\textwidth]{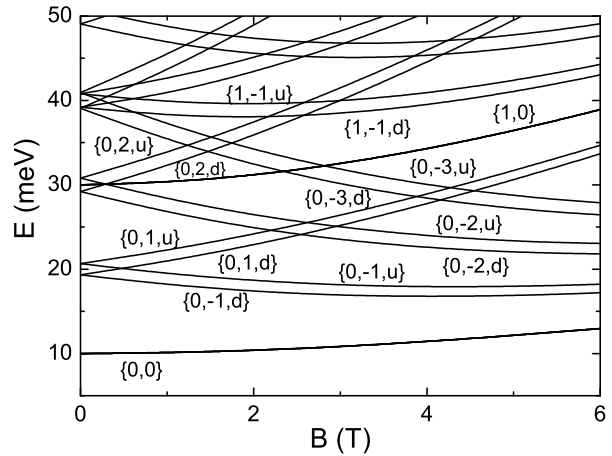}
\caption{Magnetic field dependent level scheme with $V_c$= $10$
meV at $\alpha$ = 1$\times 10^{-11}$ eV$\cdot$m. The magnetic
field induced energy gaps characterize the level assembles. The
energy levels are just Fock-Darwin states with Rashba SOI.}
\label{fig4}
\end{figure}

For case of a strongly confined dot with $V_c=10$ meV, the B-field
dependence of the energy levels are shown in Fig. \ref{fig4} at a
fixed $\alpha=10^{-11}$ eV$\cdot$m. From this figure, we see
clearly that spin-up and -down levels with the same $(N,m)$ number
are in pairs, where the energy difference between the up and down
states is determined mainly by the confinement potential and the
strength of the SOI. It can be seen further from Fig. \ref{fig4}
that in the presence of SOI, a magnetic field still plays a role
in lowering the energies of those states with negative $m$
numbers. These levels lowered by the magnetic field are known as
the Fock-Darwin~\cite{Fock,Darwin} levels and their basic features
are not affected significantly by the SOI.

The results shown in Fig. \ref{fig3} and \ref{fig4} suggest that
at relatively high-B fields so that $A_\alpha\ll 1$, the  effect
of SOI on energy spectrum of a quantum dot can be neglected. For
case of a weakly confined dot, free electron
(Landau-)\cite{Landau} type levels can be used to describe the
energy spectra of the system. When the cyclotron energy is larger
than the confining potential of a dot, there is a hybridization of
the Landau levels from the spatial confinement.

\section{Concluding Remarks}

In this paper, we have examined how the Rashba SOI affects the
energy levels of a parabolically confined quantum dot. A
non-perturbative approach to deal with SOI in a quantum dot has
been developed. The main theoretical results obtained from this
study are summarized as follows.

In the presence of a perpendicular magnetic field, we enter a
regime with different competing energies, where magnetic
potential, confining potential of the dot and potential induced by
the SOI are coupled. As a result, the presence of the magnetic
field can result in much richer features of the spintronic
properties in a quantum dot. The energy spectra and the value of
$\alpha_c$ in a dot depend strongly on the strength of the
magnetic field. In different B-field regimes, the energy levels of
spin-modified states have different dependence of the magnetic
field.

We have found that a coefficient $A_\alpha=(\alpha/\hbar)\sqrt{
2m^*/\hbar\Omega}$ with $\Omega=\sqrt{\omega_c^2+\omega_0^2}$
plays a role in switching the SOI. When $A_\alpha \gg 1$ a strong
effect of the SOI on energy levels can be observed, whereas the
effects of SOI can be neglected when $A_\alpha \ll 1$. Thus, (1)
the strong spin polarization can be achieved in a weakly confined
dot at zero and non-zero magnetic fields; (2) the characteristics
of the Fock-Darwin scheme and the Landau type levels are revealed
when the magnetic field is strong enough; and (3) the effective
SOI in the system decreases with increasing magnetic field and
confining potential of the dot.

On the basis that the energy levels in spin-degenerate quantum dot
systems have been well studied using optical and optoelectronic
measurements, we believe these experimental techniques can also be
used to examine the energy spectra of a quantum dot with SOI. We
therefore hope that the theoretical results obtained in this study
can be verified experimentally.

Although the present work deals with single-particle properties of
a quantum dot in the presence of SOI, some many-body effects
induced by Coulomb interaction can be investigated on the basis of
these results. For example, using the energy spectrum and
wavefunction obtained from this study, we can calculate the
collective excitation modes and fast-electron optical spectrum
using, e.g., a random-phase approximation~\cite{Xu}. However, for
the case of a spin-split quantum dot, these further studies
require numerical calculations considerably and we therefore do
not attempt them in the present work.

\begin{acknowledgments}
We thank T.F. Jiang and P.G. Luan for valuable discussions. W.X.
is a Research Fellow of the Australian Research Council and C.S.T.
is a Staff Scientist of the National Center for Theoretical
Sciences (NCTS) in Taiwan. This work was also supported by the
National Science Council of Taiwan under Grant No.
91-2119-M-007-004 (NCTS).
\end{acknowledgments}


\begin{thebibliography}{99}

\bibitem{Thornton} T.J. Thornton, Rep. Prog. Phys. \textbf{58}, 311
(1995).

\bibitem{Loss} D. Loss and D.P. DiVincenzo, Phys. Rev. A \textbf{57}, 120
(1998).

\bibitem{Benjamin} S. Benjamin and N.F. Johnson, Appl. Phys. Lett.
\textbf{70}, 2321 (1997).

\bibitem{Lent} C.S. Lent and P.D. Tougaw and W.Porod, Appl. Phys.
Lett. \textbf{62}, 714(1993).

\bibitem{Tougaw} P.D. Tougaw and C.S. Lent, J. Appl. Phys. \textbf{75},
1818 (1994).

\bibitem{Molotkov} S.N. Molotkov and S.S. Nazin, JEPT Lett. \textbf{63}, 687
(1996).

\bibitem{Saito} H. Saito, K. Nishi, I. Ogura, S. Sugou, and Y. Sugomito,
Appl. Phys. Lett. \textbf{69}, 3140 (1996).

\bibitem{Wolf} S.A. Wolf, D.D. Awschalom, R.A. Buhrman, J.M. Daughton,
S. von Moln\'{a}r, M.L. Roukes, A.Y. Chtchelkanova, and D.M.
Treger, Science \textbf{294}, 1488 (2001).

\bibitem{FET} B. Datta, S. Das, Appl. Phys. Lett. \textbf{56}, 665 (1990)

\bibitem{wang}  X. F. Wang, P. Vasilopoulos, and F.M. Peeters, Phys.
Rev. B \textbf{65}, 165217 (2002).

\bibitem{Tko}  T. Koga, J. Nitta, H. Takayanagi, and S. Datta, Phys. Rev.
Lett. \textbf{88}, 126601 (2002).

\bibitem{scha} Th. Sch\"apers, G. Engels, J. Lange, Th. Klocke, M.
Hollfelder, and H. L\"uth, J. Appl. Phys. \textbf{83}, 4324
(1998).

\bibitem{Rashba} Y.A. Bychkov and E.I. Rashba, J. Phys. C \textbf{17}, 6039
(1984).

\bibitem{Tarucha} S. Tarucha, D.G. Austing, T. Honda, R.J. van der Hage,
and L.P. Kouwenhoven, Phys. Rev. Lett. \textbf{77}, 3613 (1996).

\bibitem{Grundler} D. Grundler, Phys. Rev. Lett. \textbf{84}, 6074 (2000).

\bibitem{bnm} B.N. Murdin, A.R. Hollingworth, M. Kamal-Saadi, R.T. Kotitschke,
C.M. Ciesla, C.R. Pidgeon, P.C. Findlay, H.P.M. Pellemans,
C.J.G.M. Langerak, A. C. Rowe, R. A. Stradling, and E. Gornik,
Phys. Rev. B \textbf{59}, R7817 (1999).

\bibitem{Krahne} R. Krahne, V. Gudmundsson, C. Heyn, and D. Heitmann, Phys. Rev. B
\textbf{63}, 195303 (2001).

\bibitem{prl} B.I. Halperin, A. Stern, Y. Oreg, J.N.H.J. Cremers, J.A. Folk,
and C.M. Marcus, Phys. Rev. Lett. \textbf{86}, 2106 (2001).

\bibitem{Bulgakov} E.N. Bulgakov and A. F. Sadreev, JETP Lett. \textbf{73}, 505
(2001).

\bibitem{Reed} M.A. Reed, J.N. Randall, R.J. Aggarwal, R.J. Matyi, T.M. Moore,
and A. E. Wetsel, Phys. Rev. Lett. \textbf{60}, 535 (1988).

\bibitem{Governale} M. Governale, Phys. Rev. Lett. \textbf{ 89}, 206802 (2002).

\bibitem{Pfan} D. Pfannkuche, V. Gudmundsson, and P. A. Maksym,
Phys. Rev. B \textbf{ 47}, 2244 (1993).

\bibitem{Vosko01} O. Voskoboynikov, C.P. Lee, and O.
Tretyak, Phys. Rev. B \textbf{ 63}, 165306 (2001).

\bibitem{Fock} V. Fock, Z. Physik \textbf{ 47}, 446 (1928).

\bibitem{Darwin} C.G. Darwin, Proc. Cambridge Philos. Soc. \textbf{ 27}, 86
(1930).

\bibitem{Landau} L. Landau, Z. Physik \textbf{ 64}, 629 (1930).

\bibitem{Xu} For case of a spin-split 2DEG, see, e.g., W. Xu, Appl. Phys.
Lett. \textbf{82}, 724 (2003).

\end{thebibliography}
\end{document}